\documentclass[preprint,5p]{elsarticle}

\usepackage{lineno,hyperref}
\modulolinenumbers[5]

\journal{Astronomy and Computing}

\begin{document}

\begin{frontmatter}

\title{The Auroral Planetary Imaging and Spectroscopy (APIS) service}

\author[affiliation1]{L. Lamy\corref{mycorrespondingauthor}}
\author[affiliation1]{R. Prang\'e}
\author[affiliation1]{F. Henry}
\author[affiliation2]{P. Le Sidaner}

\address[affiliation1]{LESIA, Observatoire de Paris, CNRS, UPMC, Univ. Paris Diderot, 92190 Meudon, France.}
\address[affiliation2]{DIO-VO, CNRS, Observatoire de Paris, 75014, Paris, France.}
\cortext[mycorrespondingauthor]{Corresponding author}

\begin{abstract}
The Auroral Planetary Imaging and Spectroscopy (APIS) service, accessible online, provides an open and interactive access to processed auroral observations of the outer planets and their satellites. Such observations are of interest for a wide community at the interface between planetology and magnetospheric and heliospheric physics. APIS consists of (i) a high level database, built from planetary auroral observations acquired by the Hubble Space Telescope (HST) since 1997 with its mostly used Far-UltraViolet spectro-imagers, (ii) a dedicated search interface aimed at browsing efficiently this database through relevant conditional search criteria and (iii) the ability to interactively work with the data online through plotting tools developed by the Virtual Observatory (VO) community, such as Aladin and Specview. This service is VO compliant and can therefore also been queried by external search tools of the VO community. The diversity of available data and the capability to sort them out by relevant physical criteria shall in particular facilitate statistical studies, on long-term scales and/or multi-instrumental multi-spectral combined analysis.
\end{abstract}


\end{frontmatter}


\section{Introduction}

Ultraviolet (UV) planetary astronomy provides a wealth of information on planetary environments of the solar system and beyond \citep[and refs therein]{Gomez_PSS_14} and benefits a good angular resolution. Planets, moons and rings primarily reflect the (time variable) solar continuum \citep{Cessateur_AA_11} with various albedos, depending on their composition and dynamics. The UV domain is also adapted to measure intrinsic atmospheric emissions such as airglow and aurorae, which are produced by electronic transitions of neutral species prevailing in the upper atmosphere of giant planets (as H and H$_2$) and their satellites (as O) that are collisionally excited by precipitating charged particles. Airglow is a weak radiation emitted over the whole atmospheric disc and powered by solar fluorescence and photoelectron excitation \citep[and refs therein]{Barthelemy_Icarus_14}. In contrast, aurorae are bright localized emissions radiated from the auroral regions of magnetized planets (or of conductive moons interacting with them) by energetic charged particles accelerated farther in the magnetosphere \citep[and refs therein]{Badman_SSR_14}. Auroral observations, on which we focus hereafter, thus provide direct contraints on the electrodynamic interaction between planetary atmospheres, magnetospheres, moons and the solar wind as well as on the underlying plasma processes at work (particle acceleration, energy and momentum transfert). Auroral spectro-imaging in the UV provide key observables, such as the spatial topology and dynamics of the active magnetic field lines, the radiated and precipitated power, and even the energy of precipitating electrons \citep{Gustin_JMS_13,Tao_GRL_14}.

\begin{figure*}[ht!]
 \centering
 \includegraphics[width=0.8\textwidth,clip]{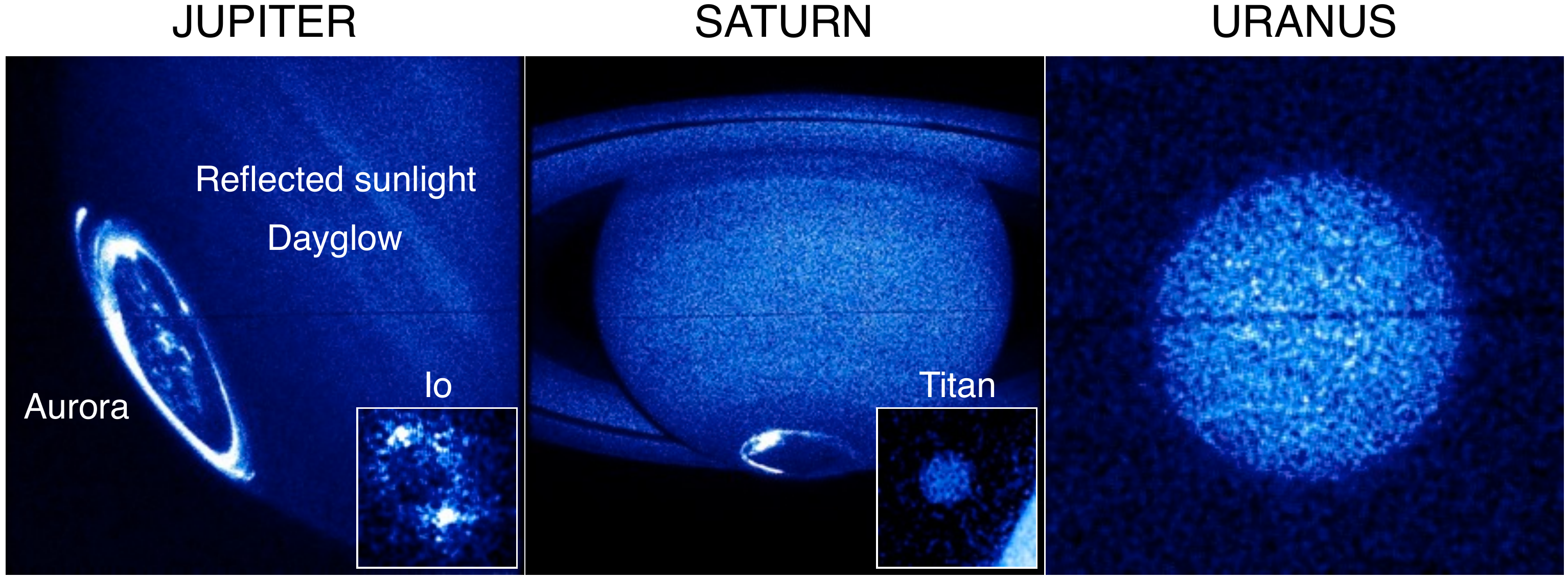}      
  \caption{Images of outer planets and satellites obtained by the ESA/NASA Hubble Space Telescope in the Far-UV, available through the APIS database.}
  \label{fig1}
\end{figure*}

Among space-based UV observatories, the Hubble Space Telescope (HST) intensively observed the aurorae of outer planetary systems (Jupiter, Saturn, Uranus) in the Far-UV (FUV, 116 to 200~nm) from 1993 up to now, providing thousands of images and spectra, often in the frame of combined observations with spatial probes dedicated to planetary exploration such as Galileo (in orbit around Jupiter over 1995-2003), Cassini (flyby of Jupiter in 2000, in orbit around Saturn since 2004) and New Horizons (flyby of Jupiter in 2007) or with Earth-based observatories observing in the radio, IR and X-rays domains. Figure~\ref{fig1} displays examples of HST-FUV observations of giant planetary systems. These observations now form a rich database, of interest for a wide community interested in planetary magnetospheres, planetology and heliospheric physics in the outer solar system. However, their use by non-specialists remains limited by their complexity, the lack of high level data and associated Figures, and the difficulty to access them. 

The Auroral Planetary Imaging and Spectroscopy (APIS) service, accessible at \url{http://apis.obspm.fr}, aims at providing an open and easy access to a high-level auroral database, built from public HST observations processed in convenient formats, and compatible with virtual observatory (VO) facilities. It thus contributes to current efforts of the VO planetology community to develop standardized databases and tools to access them. It has been open to the community in July 2013 \citep{Lamy_MOP_13}. Fortuitously, the bull APIS is also the ancient egyptian god of (data) fertilization, wearing an (active) solar disc between its horns.

\begin{figure*}[ht]
 \centering
 \includegraphics[width=1\textwidth]{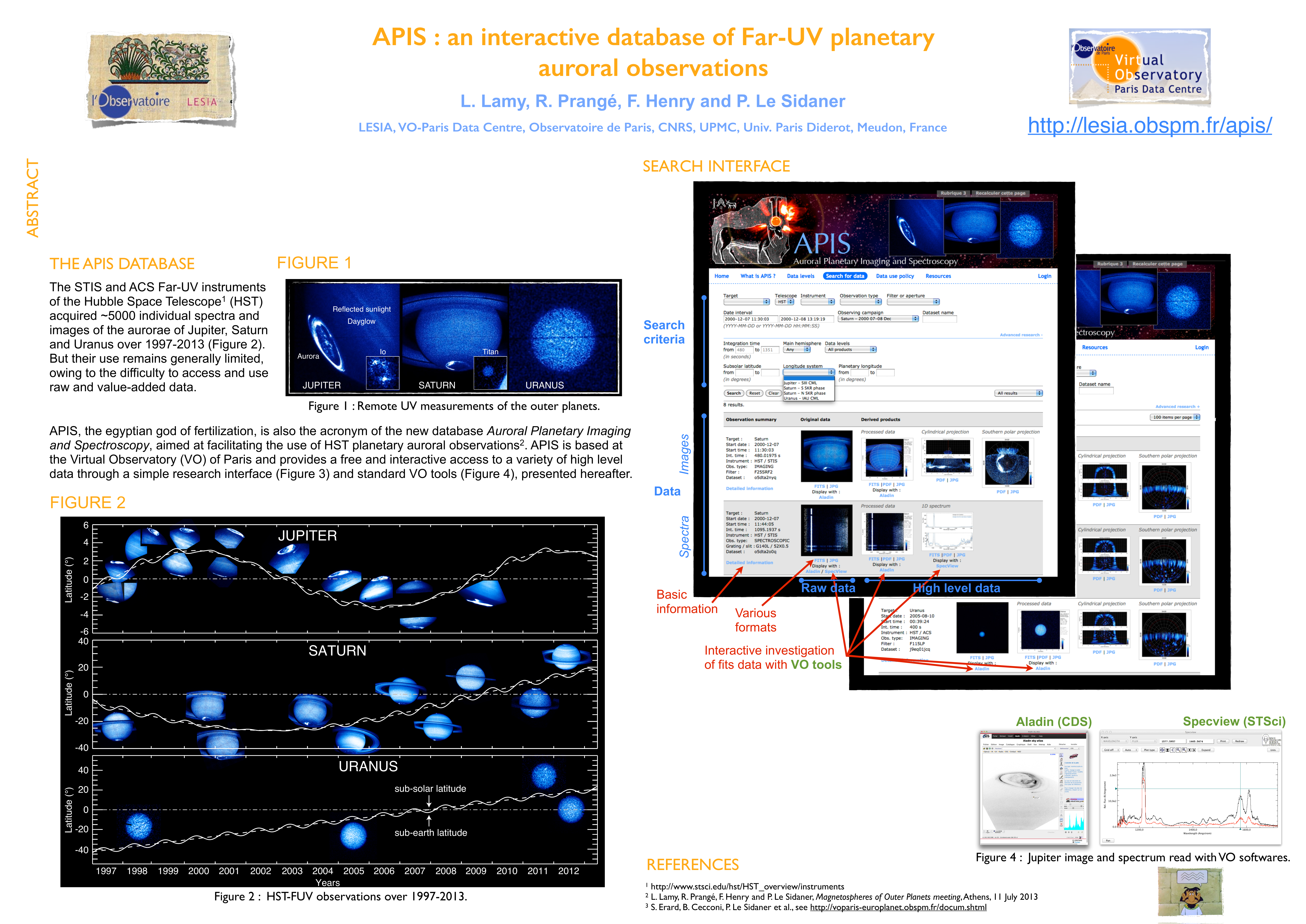}      
  \caption{Sub-Earth (solid line) and Sub-solar (dashed line) latitude of Jupiter, Saturn and Uranus as a function of time, together with combined HST images indicating the main STIS/ACS observing campaigns since 1997.}
  \label{fig2}
\end{figure*}

\section{A high level database}

APIS primarily relies on a high level database, archived at Virtual Observatory - Paris Data Centre (VO-PDC) and based on HST FUV auroral observations of the outer planets and their moons acquired since 1997. For convenience, we restricted to the mostly used instruments, namely the Space Telescope Imaging Spectrograph (STIS) and the Advanced Camera for Surveys (ACS), using the using the FUV and the Solar Blind Chanel (SBC) Multi-Anode Microchannel Array (MAMA) detectors, respectively. We refer the interested reader to the STIS and ACS handbooks provided by the Space Telescope Science Institute (STScI) for further details. To date, the considered observations consist of about 6000 individual images and spectra, issued from 37 observational campaigns of the Jupiter, Saturn and Uranus systems distributed as illustrated in Figure \ref{fig2}, and obtained with very diverse instrumental configurations (filters for imaging, combination of slit and gratings for spectroscopy). For each of these individual observations, APIS provides a set of value-added data levels briefly described below, each available in work and graphical formats such as {\it fits} files \citep{Pence_AA_10} and {\it jpeg, pdf} plots. 

\begin{figure*}[ht!]
 \centering
 \includegraphics[width=1\textwidth,clip]{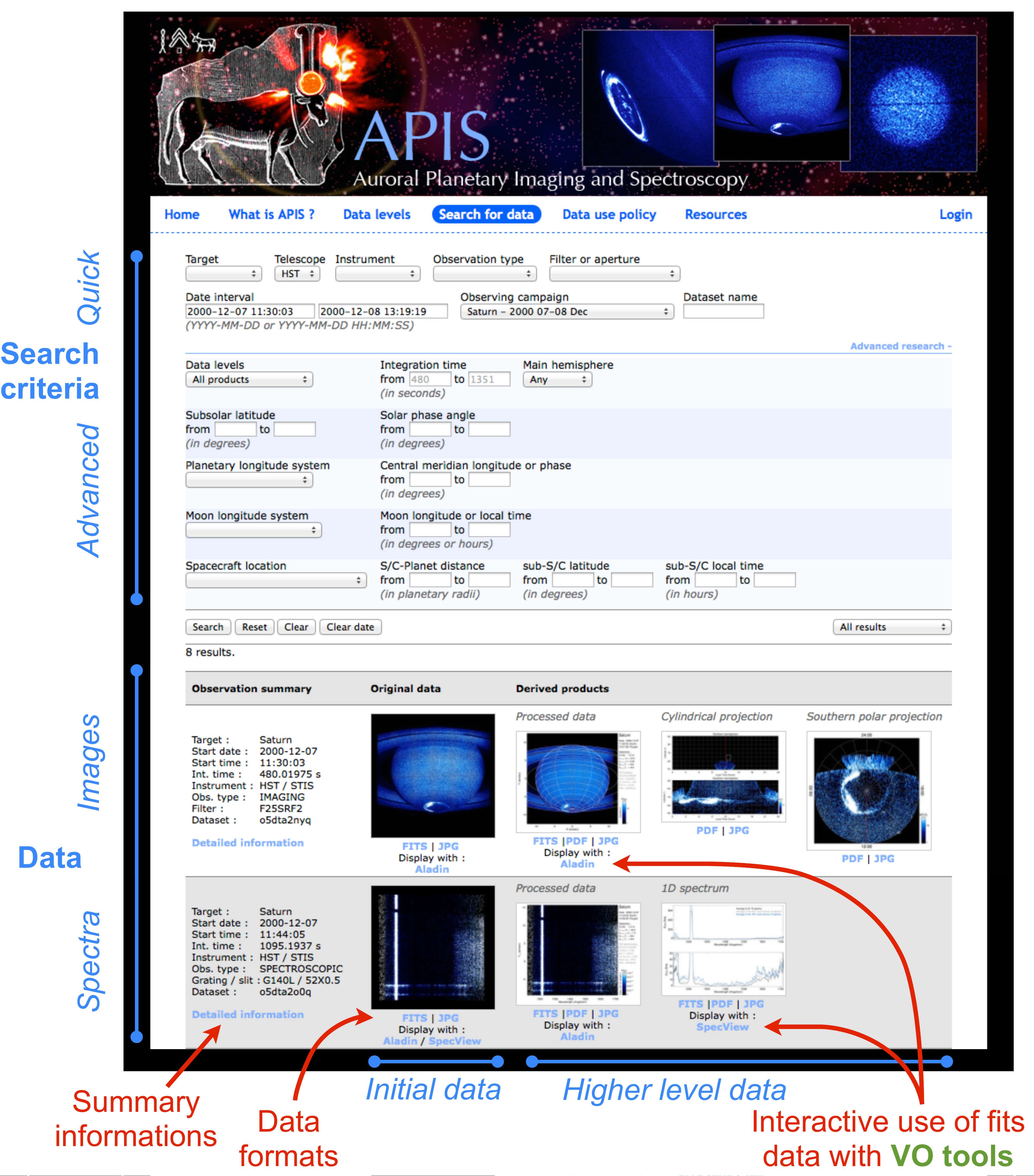}      
  \caption{Search interface.}
  \label{fig3}
\end{figure*}

\subsection{Imaging}

Images are provided under three levels of data : original images as calibrated by STScI (level 1), spatially re-oriented and fitted images provided with pixel planetocentric coordinates (level 2) and background-subtracted cylindrical/polar projections transposed into physical units (level 3). At the bottom of Figure \ref{fig3}, the row labelled "Images" (in blue) illustrates these data levels with quick looks, below which are provided various data formats.

\paragraph{Level 1}

Original images - or level 1 images - originate from public data delivered by STScI through the Mikulski Archive for Space Telescopes (MAST), and mirrored at the ESA Hubble Science Archive. They correspond to calibrated images in the sense intended by STScI, which means that they have undergone a standard processing (correction for dark current, flat-fielding, correction for geometric distortion) and are expressed in basic units, such as counts.pix$^{-1}$  for STIS and electrons.s$^{-1}$.pix$^{-1}$ for ACS, with the X and Y axis displayed in the telescope geometric frame. 

For our purpose, we have systematically reprocessed the STIS calibrated images ({\it x2d} extension) from raw data through the standard STScI pipeline with the most recent calibration files, which are referenced in the header of updated {\it fits} files. Alternately, we used the calibrated ACS {\it fits} files ({\it drz} extension) directly as retrieved from MAST, as these are calibrated up-to-date by default. Before further processing, the (dark) line of bad pixels visible in ACS images was replaced by values linearly interpolated from neighbor pixels.

Plots of level 1 images display the telescope field-of-view (FOV) in the telescope frame with a linear intensity scale. 

\paragraph{Level 2}

Processed images - level 2 - are obtained as follows. Level 1 images are first rotated to be displayed in a target-related physical frame. The Y axis points along the rotation axis for all bodies, except for Uranus where it points along the celestial north, and the X axis points Eastward. 

The images are then individually corrected from the HST pointing inaccuracy, which can reach a few tens of pixels in translation and a few fractions of a degree in rotation. In practice, this essential step is achieved by fitting the planetary limb at the 1-bar level, as well as the rings when visible, by a reference ellipsoid. This is done automatically through 2D cross-correlation for small size targets (Uranus and moons) and ACS images of Saturn, with an accuracy estimated to 1 pixel in both X and Y directions. As STIS observations of Saturn display a noisier limb which can hardly be automatically fitted, the corresponding images (about 70 in total) were all individually re-centered manually, with an accuracy estimated to 1-2 pixels. The more complex fitting of Jupiter images, for which only part of the planetary disc is visible, is under progress. 

The level 2 images are then extracted from the re-centered images with a square box centered on the target and whose size varies with that of the observed body.

\begin{figure}[ht!]
 \centering
 \includegraphics[width=0.5\textwidth,clip]{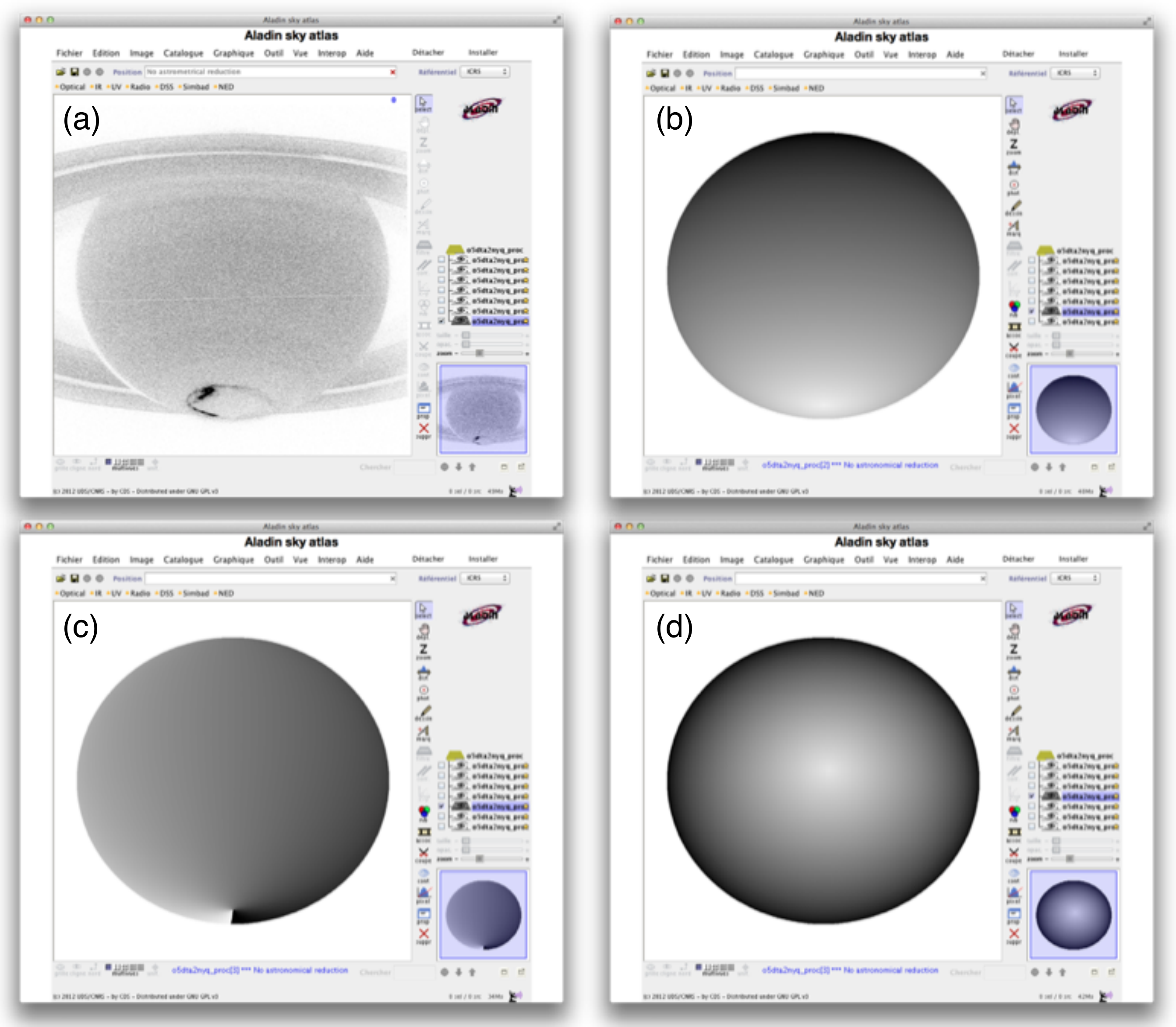}      
  \caption{Example of Aladin's use with a level 2 image of Saturn. Panels (a), (b), (c) and (d) respectively plots the extensions 0 (science image), 1 (latitude at the limb altitude), 2 (local time at the limb altitude) and 3 (solar zenithal angle at the limb altitude) of the level 2 {\it fits} file. The plotted extension is highlighted in blue on the right bar.}
  \label{fig4}
\end{figure}

The {\it fits} files provided for level 2 images contain 7 science extensions together with an updated header. Extension 0 corresponds to the re-centered image in unchanged units. Extensions 1 to 6 provide the associated pixels coordinates in a planetocentric frame, namely the latitude (extension 1), the local time (extension 2), the zenithal solar angle (extension 3) and the zenithal observing angle (extension 4) at the limb altitude, together with the latitude (extension 5) and the local time (extension 6) at the auroral peak altitude, hence at the footprint of active magnetic field lines. Figure \ref{fig4} shows plots of extensions 0 to 3 for a level 2 image of Saturn. The header contains basic information on the original observation together with value-added information as the planetary ephemeris (source : Institut de M\'ecanique C\'eleste et de Calcul des Eph\'em\'erides or IMCCE, \url{http://www.imcce.fr}), the ephemeris of moons and Galileo, Cassini spacecraft for Jupiter and Saturn (source : University of Iowa, \url{http://www-pw.physics.uiowa.edu}, based on SPICE kernels), the Saturn's Kilometric Radiation (SKR) southern and northern phases \citep{Lamy_PRE7_11} (source : Cassini/RPWS/HFR Kronos server, \url{http://www.lesia.obspm.fr/kronos/skr_periodicity.php}), and resolved secondary targets coincidently visible in the HST FOV (such as Callisto for Jupiter, Titan for Saturn or Saturn for Titan).

Plots of level 2 images display the telescope FOV in the target frame with a logarithmic intensity scale, aimed at reducing the contrast between bright and faint emissions, and a white grid of planetocentric latitude and local time at the limb altitude. The red meridian marks noon. Basic information on the observation and the associated ephemeris are reminded on the right.

Level 2 data are intended to be useful to the broadest range of users, as the latter have the possibility to take advantage of pixel coordinates (latitude, longitude, zenithal angles) to derived and remove any background model of their choice and/or to apply any preferred conversion factor to transpose basic units to physical ones.

\paragraph{Level 3}

Cylindrical and polar projections at auroral altitudes expressed in physical units - level 3 - are finally derived for the images of planets through the following additional processing steps. 

\begin{figure*}[ht!]
 \centering
 \includegraphics[width=0.9\textwidth,clip]{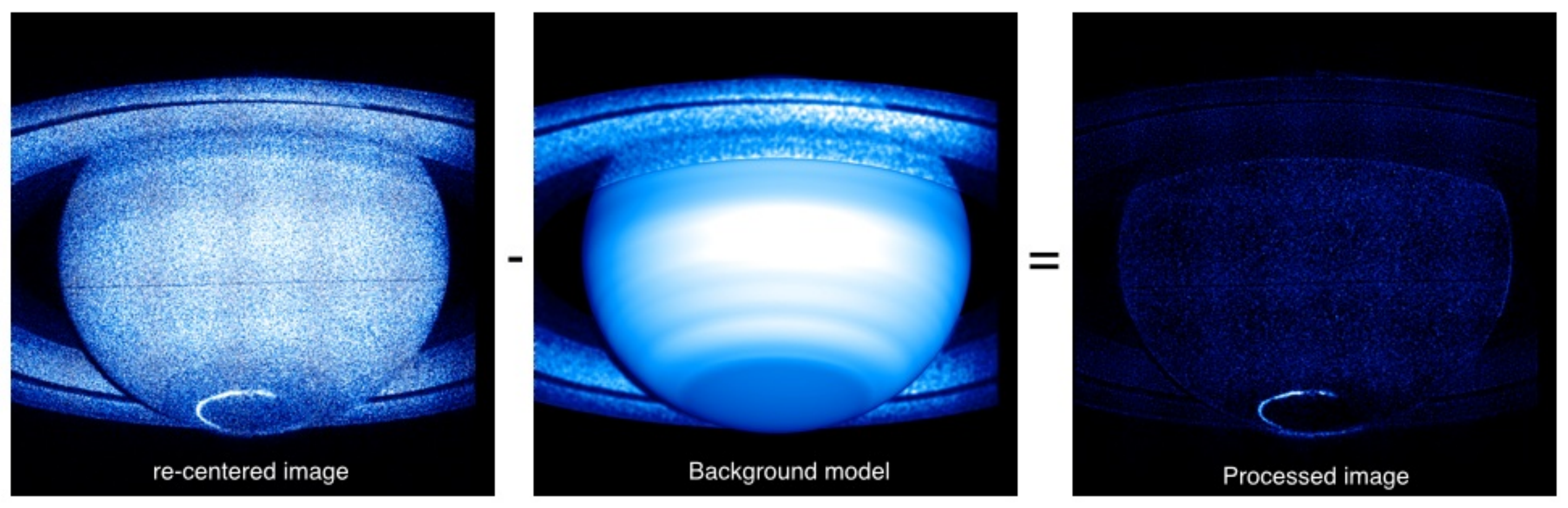}      
  \caption{Example of subtraction of a numerical background disc model, here completed by an empirical ring background model, to a STIS image of Saturn.}
\label{fig3bis}
\end{figure*}

A model of reflected background at the limb altitude is first subtracted to level 2 images. Several types of backgrounds have been employed and tested in the literature. The most accurate solution is generally to derive an empirical background from one or several images without auroral signal acquired close to the investigated observation. As this is not always possible to find relevant references images for a given campaign and because the amount of observations requires an automated processing pipeline, we rather chose to subtract a Minnaert-type numerical background model \citep[and refs therein]{Vincent_Icarus_00} fitted to the non-auroral regions of the observed planetary disc and then convolved by the STIS or ACS point spread function. The Minnaert model consists of a polynomial law of the neperian logarithm of the product of the cosines of the observation and solar zenith angles. We tested different sets of polynomial orders and latitudinal bins separately for each planet. The best results were obtained with first order polynoms and latitudinal bins of 2$^\circ$ and 30$^\circ$ for Saturn and Uranus respectively. In the case of Jupiter, we note that a first order Minnaert polynom with latitudinal bins of 1$^\circ$ has been used \citep{Bonfond_10}. The uncertainty of the model background in the auroral regions was estimated for Saturn to be less than 10$\%$, and less for Uranus. For the sake of completeness, we also tested an alternate numerical background model, developed for FUV auroral images in the case of the Earth and based on the sum of the cosines of the observation and solar zenith angles \citep{Li_IEEE_04}. This did not provide background models of higher accuracy though. An example of the applied processing is illustrated with a STIS image of Saturn on Figure \ref{fig3bis}.

Once the background subtracted, the resulting images are converted into auroral brightnesses (in units of kilo-rayleighs or kR, with 1~R~=~10$^6$ photons.s$^{-1}$.cm$^{-2}$ in 4$\pi$ sr) of total H$_2$ emission over the 80-170nm spectral range. Many past studies have performed successive photometric calibrations of STIS and ACS with typical auroral spectra of planets to convert instrumental units into auroral brightnesses \citep{Gerard_JGR_04,Gerard_JGR_06,Grodent_JGR_05,Clarke_JGR_09,Lamy_08,Bonfond_10}. The resulting conversion factors yielded significant disparities, owing to the considered radiative species (H$_2$ and/or H-Ly$\alpha$), the consideration of absorption by hydrocarbons and the selected bandpass (spectral domain of H$_2$ emission, FUV domain or bandpass restricted to that of HST imaging filters), as discussed by \citep{Gustin_JGR_12, Lamy_JGR_13}. Here, we used the most recent estimates of \cite{Gustin_JGR_12} derived for the two broadband filters of each instrument, which are the clear and SRF2 filters for STIS and the F115LP and F125LP ones for ACS, that we extrapolated to the other employed filters. The conversion factors of filters outside of the spectral domain of auroral H$_2$, as F165LP for ACS, were set to 0. 

The obtained images are finally projected at the auroral peak altitude, indicated in the header of level 2 {\it fits} files, with a dedicated projection routine preserving the photon flux of a given emitting surface both in cylindrical and polar frames. Cylindrical and polar projections are systematically derived for both northern and southern hemispheres, but the search interface (described in the next section) only displays the projection of the main (summer) hemisphere. The projection of the secondary (winter) hemisphere, together with all data levels are accessible when clicking on the "Detailed information" link, visible on the left of each observation in Figure \ref{fig3}. Such projections were derived for planets only, and not for moons, owing to the different radiative species and the lack of reliable conversion factors.

Plots of level 3 images display cylindrical and polar projections with a logarithmic intensity scale and a white grid of planetocentric latitude and local time at the peak auroral altitude. The red meridian marks noon.

To help the APIS user to track auroral dynamics, several visual references were additionally added to Saturn's projections. Cylindrical projections display white boxes in both hemispheres at the southern and northern expected footprints of Enceladus. The latter was indeed detected by Cassini/UVIS \citep{Pryor_Nature_11}, but never identified in HST images so far \citep{Wannawichian_JGR_08}. Kronian cylindrical and polar projections also display dashed-dotted (dashed) lines to mark the southern (northern) reference rotating meridian derived from the southern (northern) SKR phase system \citep{Lamy_JGR_13}.

\subsection{Spectroscopy}

Spectra included in the APIS database so far restrict to first-order long-slit, or slitless, spectroscopic measurements acquired with STIS. They are also provided under three levels of data : original 2D spectra as calibrated by STScI (level 1), extracted 2D spectra which are re-calibrated in wavelength whenever needed and possible (level 2) and 1D average spectra built from different occurrence levels (level 3). At the bottom of Figure \ref{fig3}, the row labelled "Spectra" (in blue) illustrates these data levels with quick looks and the available data formats.

\paragraph{Level 1}

As for images, original 2D spectra - level 1 - are retrieved from the MAST archive and individually re-calibrated ({\it x2d} extension) along the STScI processing pipeline with the most recent calibration files. They are expressed in physical units, namely erg.s$^{-1}$.cm$^{-2}$ .angstrom$^{-1}$.arcsec$^{-2}$ by default, and displayed in the telescope geometric frame such that the X axis corresponds to wavelengths and the Y axis corresponds to the spatial direction along the slit (or, equivalently, the Y axis of level 1 images).

Plots of level 1 spectra display 2D spectra with a linear intensity scale, saturated at H-Ly$\alpha$. As planetary targets are extended, the width of the H-Ly$\alpha$ line provides the effective spectral resolution and, when the target fully fills in the slit, the width of the H-Ly$\alpha$ line corresponds to that of the slit.

\paragraph{Level 2}

Processed 2D spectra - level 2 - are extracted from original 2D spectra, and re-calibrated in wavelength when not already performed by STScI. In practice, the latter step was done by fitting the H-Ly$\alpha$ line whenever contained in the observed spectral range. 

The {\it fits} files provided for level 2 spectra contain a single science extension, corresponding to the derived 2D spectrum, together with an updated header similar to that of level 2 images, which mentions the used spectral calibration (STScI, APIS or none).

Plots of level 2 spectra are displayed with a linear intensity scale and axis coordinates. Basic information on the observation and the associated ephemeris are displayed on the right.

\paragraph{Level 3}

As 2D spectra are not necessarily a convenient format to deal with, a set of typical 1D spectra - level 3 - are provided for each long-slit spectral observation. They consist of three average spectra respectively derived from the 50\%, 10\% and \%1 most intense 1D spectra (which are compared once integrated in wavelength) acquired along the slit. These average spectra thus provide an indication of the mean, high and peak spectral activity during the observation, regardless of the target size though. They have been expressed in R.angstrom$^{-1}$ to facilitate comparisons with the literature.

The {\it fits} files provided for the level 3 spectra contain three science extensions, corresponding to the three 1D average spectra described above, together with an updated header similar to that of level 2 spectra.

Plots of level 3 spectra display the 1D spectra smoothed over 5 pixels with a linear intensity scale in two panels. The intensity range of the upper panel is variable, in order to show the profile of the most intense signal (generally the H-Ly$\alpha$ line), while the intensity range of the lower panel if fixed, to facilitate comparisons between weaker emissions as those of H$_2$.

\section{Search interface}


The above database can be queried by a dedicated search interface available on the APIS web portal, as displayed on Figure \ref{fig3}. A data query consists of a conditional search over a set of parameters (such as the date, the target, the instrument, the observation name etc.), known as metadata which describe each single observation.

\subsection{Quick search}

The selection criteria located right to the blue label "Quick" on Figure \ref{fig3} are always visible. They consist of basic criteria needed to perform a quick search. The first row lists fields dealing with instrumental specifications : Target (name of the Planet or Satellite), Telescope (HST by default), Instrument (ACS or STIS), Observation type (Imaging or Spectroscopy), Filter or Aperture (filters for imaging, any combination of slit and gratings for spectroscopy), the second row lists fields dealing with temporal selection : Data Interval, Observing Campaign (Data grouped by HST observing program) and Dataset (unique HST data identifier).

An example of data request (selection of the observing campaign {\it Saturn - 2000 07-08 Dec}) is illustrated at the bottom of the Figure. The data are sorted out by time vertically and the different data levels appear in successive horizontal columns. The first column, entitled "Observation summary", provides a list of basic parameters associated to each observation. More parameters are displayed by clicking in on the "Detailed information" link at the bottom of each list. For each data level, the available data formats lie below a quick look. Clicking in on the latter displays the full resolution plot and enables the user to pass from one observation to the following or preceding one for a chosen data level.

\subsection{Advanced search}

The selection criteria located right to the blue label "Advanced" on Figure \ref{fig3} automatically appear when clicking in on the "Advanced Search" button, at the bottom right of the standard quick search interface. These supplementary criteria have been chosen according to the typical needs of the community. The available fields divide in five rows : (i) Data level, range of Integration time and Main hemisphere (North or South), (ii) ranges of Sub-solar latitude and of Solar phase angle, (iii) Planetary longitude system (System III for Jupiter, Northern or Southern SKR phase system for Saturn, IAU longitude system for Uranus) and range of Central meridian longitude (CML) or Phase, (iv) Moon longitude system (Local Time or Longitude for the main satellites of Jupiter and Saturn) and range of Local Time/Longitude and (v) Spacecraft Location (Galileo for Jupiter, Cassini for Saturn) and ranges of Distance, Latitude and Local time relative to the host planet.

To facilitate iterative searches, the ranges of fields Time interval and Integration time are automatically filled in when performing a request. These parameters also appear by default for each observation among the summary information displayed in the left column of Figure \ref{fig3}. When performing a conditional search on parameters not present among these (say, for instance, the Local Time of Enceladus), they are automatically added to the list of summary information.

\subsection{Science uses}

In the past, HST auroral observations have been most generally analyzed per observing campaign. The above described database and search interface further enable the non-specialist, but also the specialist user, to easily find individual observations of interest over a larger set of observing campaigns as a function of an extended space of parameters : the sub-solar latitude to look for seasonal effects (Figure \ref{fig2}), the satellite position to search for planet-satellite electrodynamic interactions, the planetary longitude/phase to investigate any rotational control on magnetospheric dynamics, the date of arrival of interplanetary shocks to study the influence of solar wind on auroral forcing or the location of a spacecraft performing remote/in situ measurements to lead multi-instrumental multi-spectral analysis. Furthermore, with standardized data levels, APIS is particularly adapted to statistical studies, over long-term scales, and the interval between 1997 and 2014 now exceeds the duration of a solar cycle, a jovian revolution or half a kronian revolution around the sun.

\section{VO compatibility}

\subsection{Interactive use}

A last functionality of the APIS service is that the user can work with the data directly online, without needing to download them. This is achieved by using the VO softwares Aladin (\url{http://aladin.u-strasbg.fr}, developed by the Centre de Donn\'ees astronomiques de Strasbourg or CDS) for images and Specview (\url{http://www.STScI.edu/institute/software_hardware/specview/}, developed by STScI) for spectra. The user just has to click on the "Aladin" or "Specview" links visible under each data level providing a {\it fits} file. Tutorials for using APIS in general, and these VO tools in particular, are available on the web page (written document, movie).

\paragraph{Images}

Aladin is capable of reading any 2D image or spectrum, and enables the user to apply simple operations, such as plotting the various extensions of a {\it fits} file, changing the contrast, mapping intensity levels, drawing an histogram of intensities, plotting a cut of the image, identifying the intensity and coordinates of pixels of interest etc. The user does not have to download Aladin before to use it, as APIS automatically launches a JAVA applet version of the software, which opens in a new window. The data are then automatically sent toward Aladin through a Simple Application Messaging Protocol (or SAMP) hub. Further documentation on SAMP is provided by the International Virtual Observatory Alliance (IVOA) at \url{http://www.ivoa.net/documents/SAMP/}. Figure \ref{fig4} illustrates the use of Aladin with plots of several extensions of the level 2 image visible in Figure \ref{fig3}. It is also possible to upload several {\it fits} files together to compare them.

\paragraph{Spectra}

Specview is capable of reading both 2D and 1D spectra to display a 1D plot of intensity as a function of wavelength. It also enables the user to apply simple operations, such as coplotting different spectra, make some simple measurements or add theoretical transitions lines retrieved from interoperable spectroscopy catalogs issued from the Virtual Atomic and Molecular Data Center (VAMDC) program (\url{http://www.vamdc.eu/}). By waiting for an operational JAVA applet, the user has to download the latest version of Specview and launch it before clicking on the "Specview" link below the {\it fits} file. The data are then sent to the software through a SAMP hub.

\begin{figure}[ht!]
 \centering
 \includegraphics[width=0.5\textwidth,clip]{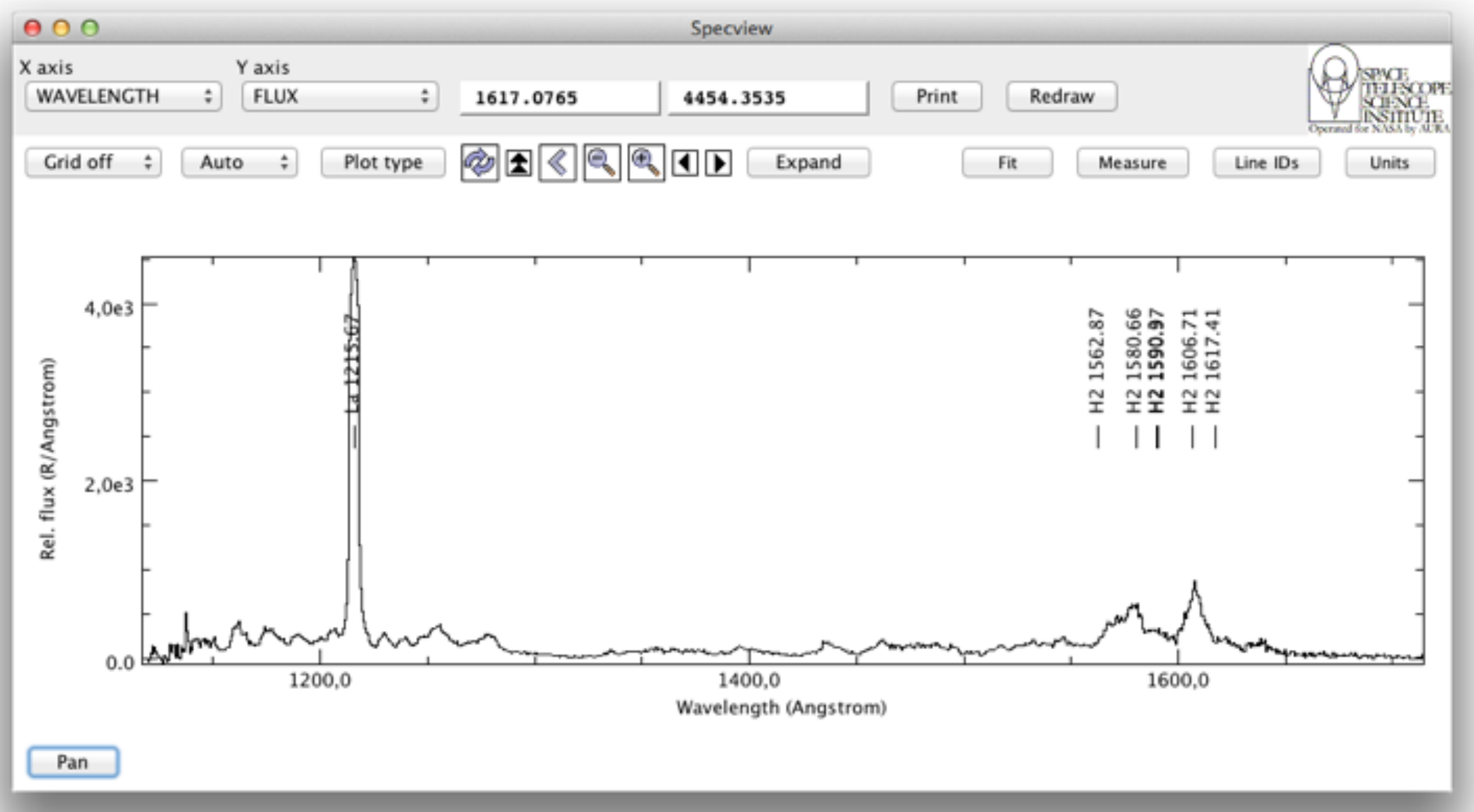}      
  \caption{Example of 1D spectrum plotted with Specview. Several theoretical lines have been retrieved and plotted on the top of the investigated spectrum through the button "Lines IDs" : the H-Ly$\alpha$ line is provided by default with Specview, while the H$_2$ lines have been retrieved from the SESAM catalog.}
  \label{fig5}
\end{figure}

When reading a 2D spectrum, the user has to choose which spatial pixels of the slit will be summed up to derive a single 1D spectrum. Figure \ref{fig5} displays a Specview plot of a 1D spectrum extracted from the level 1 spectrum visible in Figure \ref{fig3}, which reveals a typical spectrum of Saturn's auroral emissions. The button "Line IDs" enables the user to retrieve theoretical lines from a set of VAMDC catalogs, such as SESAM (\url{http://sesam.obspm.fr}) for H$_2$ lines. For instance, we superimposed the H-Ly$\alpha$ line (provided by default with Specview) and a few H$_2$ lines retrieved from the SESAM catalog to the observed spectrum.

\subsection{Interoperability}

Finally, out of the search interface presented above, the APIS database can also be queried externally, by wider search tools of the Virtual Observatory community such as TOPCAT (Tool for Operations on Tables and Catalogs, \url{http://www.star.bris.ac.uk/~mbt/topcat/}), VESPA (Virtual European Solar and Planetary Data Access, \url{http://vespa.obspm.fr}) or the CDPP/AMDA tool (Centre de Donn\'ees de Physique des Plasmas/Automated Multi Dataset Analysis, \url{http://amda.cdpp.eu/}). 

To achieve such a VO interoperability, APIS has been built to be compliant with the Europlanet - Table Access protocol (EPN-TAP), whose aim is to enable a user to access Planetary Science data in a standard manner. The EPN-TAP service relies on IVOA TAP specifications and a data model developed by the EuroPlaNet team. For details, the reader is referred to the description of the EPN-TAP protocol  \citep{Erard_AC_14}. In practice, the set of metadata attributed to each individual observation includes the (mandatory and optional) parameters requested by the EPN-TAP data model to fully describe the data. These are divided in three categories : axis ranges, data description, and data access. The metadata are organized into a single table (or view), which can be then queried using the TAP protocol between a user TAP client and the APIS TAP server.

The APIS EPN view has been officially registered by IVOA through VO-PDC. APIS was one of the first services connected to VESPA through EPN-TAP \citep{Erard_AC_14b}. A use-case of the CDPP/AMDA tool querying APIS data through EPN-TAP and a SAMP hub (as described above) for investigating the influence of solar wind conditions on kronian auroral emissions is presented in \citep{Genot_AC_14}. 


\section{Conclusions and perspectives}

In this article, we have presented in detail the APIS service. This primarily consists of a high level database, made of several data levels derived for most of the HST-FUV STIS and ACS spectro-imaging auroral observations acquired since 1997, with very diverse instrumental configurations. A dedicated search interface accessible on the APIS portal enables the user to efficiently browse and sort out these data, which can be used interactively with VO softwares such as Aladin and Specview. The database is also compliant with the EPN-TAP protocol, so that it can also be queried by external VO tools. We have presented a few examples of science studies easy to perform with this service.

This service is in continuous evolution, with numerous perspectives of development and improvement. First of all, although most of the data levels are already available for the considered targets, the level 2,3 images of Jupiter remain for instance to be derived (under progress). The data levels themselves may also be updated to improve the existing plots (for instance by drawing the magnetic footprints of Galileo and Cassini spacecraft, or the expected footprint of Galilean satellites, on top of projections) or to provide additional observables (such as the total auroral power radiated per image) and formats (such as movies). The date of creation of each dataset being contained within the header of associated {\it fits} files and displayed in the "Detailed information" link of the search interface, updates of existing data levels can be traced by the user. This service shall also evolve as a function of the feedback of the  community. It has been and will be regularly presented at science and hands-on sessions of international meetings to facilitate exchanges with the users and probe them on desired further improvements.

The database will then naturally be extended with the observing campaigns of the Jupiter, Saturn and Uranus systems to come by the end of the HST mission. But it could also be extended to the other planets observed by STIS and ACS (such as Mars), and to the other HST FUV spectro-imagers (such as the FOC and WFPC2, a few images of which were already included to APIS as a test). Comments or suggestions of improvement from the users on these possible extensions are also welcome. In parallel, simple (blind) usage statistics can be used to assess whether, and at which degree, APIS responds to the needs of the community. 

Furthermore, now that this service is fully operational, it could easily deal with much wider datasets such as the auroral spectro-imaging observations acquired by the instruments onboard the mission Cassini, or by similar spectro-imagers onboard the various past and future planetary exploration spacecraft (such as Voyager, Galileo, Polar, Mars Express, JUNO, JUICE) or Earth-based observatories (such as Chandra, XMM-Newton, IR ground-based telescopes, JWST). Another axis of development is to integrate databases of observed FUV solar spectra and/or theoretical spectra of H$_2$ bands (the SESAM catalog only provides individual lines so far). Finally, this service could be easily interfaced with other VO tools (such as HELIO, or the 3Dview and propagation tools at CDPP).

\section{acknowledgements}
We thank CNRS and CNES for supporting the authors. We acknowledge all the persons involved in the acquisition of data by the ESA/NASA Hubble Space Telescope (from the conceptors of the telescope and its instruments to the principal investigators of observing programs), the Space Telescope Science Institute (STSci) and the ESA Hubble Science Archive (ESAC) for calibrating and archiving raw data and developing interactive tools for spectra (Specview, VOspec), the IMCCE for providing the physical ephemeris of the outer planets and their moons, and the Cassini, Galileo, and Voyager ephemeris tool made available by the Radio and Plasma Wave Group from the Department of Physics and Astronomy of the University of Iowa. We acknowledge S.~Cnudde and SIGAL for designing the web page, and S.~Erard, B.~Cecconi, J.~Aboudarham and X.~Bonnin for useful discussions on the APIS development. We thank N.~Moreau and I.~Busko for developing a version of Specview compatible with APIS products, and E.~Roueff and H.~Abgrall for exchanges on the SESAM database.
\section{Bibliography styles}

\section*{References}

\bibliography{mybibfile}

\begin{thebibliography}{10}
\expandafter\ifx\csname url\endcsname\relax
  \def\url#1{\texttt{#1}}\fi
\expandafter\ifx\csname urlprefix\endcsname\relax\def\urlprefix{URL }\fi
\expandafter\ifx\csname href\endcsname\relax
  \def\href#1#2{#2} \def\path#1{#1}\fi

\bibitem{Gomez_PSS_14}
A.~I. {Gomez de Castro}, T.~{Appourchaux}, M.~A. {Barstow}, M.~{Barthelemy},
  F.~{Baudin}, S.~{Benetti}, P.~{Blay}, N.~{Brosch}, E.~{Bunce}, D.~{de
  Martillo}, J.-M. {Deharveng}, R.~{Ferlet}, K.~{France}, M.~{Garc'a},
  B.~{Gansicke}, C.~{Gry}, L.~{Hillenbrand}, E.~{Josselin}, C.~{Kehrig},
  L.~{Lamy}, J.~{Lapington}, A.~{Lecavelier des Etangs}, F.~{LePetit},
  J.~{Lopez-Santiago}, B.~{Milliard}, R.~{Monier}, G.~{Naletto}, Y.~{Naze},
  C.~{Neiner}, J.~{Nichols}, M.~{Orio}, I.~{Pagano}, C.~{Peroux}, G.~{Rauw},
  S.~{Shore}, M.~{Spaans}, G.~{Tovmassian}, A.~{ud-Doula}, J.~{Vilchez},
  {Building galaxies, stars, planets and the ingredients for life between the
  stars. The science behind the European Ultraviolet-Visible Observatory.},
  Astrophys. and Sp. Sci.

\bibitem{Cessateur_AA_11}
G.~{Cessateur}, T.~{Dudok de Wit}, M.~{Kretzschmar}, J.~{Lilensten}, J.-F.
  {Hochedez}, M.~{Snow}, {Monitoring the solar UV irradiance spectrum from the
  observation of a few passbands}, Astron. \& Astrophys. 528 (2011) A68.
\newblock \href {http://dx.doi.org/10.1051/0004-6361/201015903}
  {\path{doi:10.1051/0004-6361/201015903}}.

\bibitem{Barthelemy_Icarus_14}
M.~{Barth{\'e}lemy}, L.~{Lamy}, H.~{Menager}, M.~{Schulik}, D.~{Bernard},
  H.~{Abgrall}, E.~{Roueff}, G.~{Cessateur}, R.~{Prange}, J.~{Lilensten},
  {Dayglow and auroral emissions of Uranus in H$_{2}$ FUV bands}, Icarus 239
  (2014) 160--167.
\newblock \href {http://dx.doi.org/10.1016/j.icarus.2014.05.035}
  {\path{doi:10.1016/j.icarus.2014.05.035}}.

\bibitem{Badman_SSR_14}
S.~V. {Badman}, G.~{Branduardi-Raymont}, M.~{Galand}, S.~L.~G. {Hess},
  N.~{Krupp}, L.~{Lamy}, H.~{Melin}, C.~{Tao}, {Auroral Processes at the Giant
  Planets: Energy Deposition, Emission Mechanisms, Morphology and Spectra}, Sp.
  Sci. Rev.\href {http://dx.doi.org/10.1007/s11214-014-0042-x}
  {\path{doi:10.1007/s11214-014-0042-x}}.

\bibitem{Gustin_JMS_13}
J.~{Gustin}, J.-C. {G{\'e}rard}, D.~{Grodent}, R.~G. {Gladstone}, J.~T.
  {Clarke}, W.~R. {Pryor}, V.~{Dols}, B.~{Bonfond}, A.~{Radioti}, L.~{Lamy},
  J.~M. {Ajello}, {Effects of methane on giant planetÕs UV emissions and
  implications for the auroral characteristics}, Journal of Molecular
  Spectroscopy) 291 (2013) 108--117.
\newblock \href {http://dx.doi.org/10.1016/j.jms.2013.03.010}
  {\path{doi:10.1016/j.jms.2013.03.010}}.

\bibitem{Tao_GRL_14}
C.~{Tao}, L.~{Lamy}, R.~{Prang{\'e}}, {The H-Lyman alpha/H2 bands brightness
  ratio of FUV auroral emissions: a diagnosis for the energy of precipitating
  electrons and associated magnetospheric acceleration processes applied to
  Saturn}, Geophysical Research Letters (2014) in press.

\bibitem{Lamy_MOP_13}
L.~{Lamy}, R.~{Prang{\'e}}, F.~{Henry}, P.~{Le Sidaner}, {APIS, an interactive
  database of HST-UV observations of the outer planets}, {Presented at the
  Magnetospheres of Outer Planets conference, Athens, 11 July}, 2013.

\bibitem{Pence_AA_10}
W.~D. {Pence}, L.~{Chiappetti}, C.~G. {Page}, R.~A. {Shaw}, E.~{Stobie},
  {Definition of the Flexible Image Transport System (FITS), version 3.0},
  Astronomy \& Astrophysics 524 (2010) A42.
\newblock \href {http://dx.doi.org/10.1051/0004-6361/201015362}
  {\path{doi:10.1051/0004-6361/201015362}}.

\bibitem{Lamy_PRE7_11}
L.~{Lamy}, {Variability of southern and northern periodicities of Saturn
  Kilometric Radiation}, Planetary, Solar and Heliospheric Radio Emissions (PRE
  VII) (2011) 38--50\href {http://arxiv.org/abs/1102.3099}
  {\path{arXiv:1102.3099}}.

\bibitem{Vincent_Icarus_00}
M.~B. {Vincent}, J.~T. {Clarke}, G.~E. {Ballester}, W.~M. {Harris}, R.~A.
  {West}, J.~T. {Trauger}, R.~W. {Evans}, K.~R. {Stapelfeldt}, D.~{Crisp},
  C.~J. {Burrows}, J.~S. {Gallagher}, R.~E. {Griffiths}, J.~{Jeff Hester},
  J.~G. {Hoessel}, J.~A. {Holtzman}, J.~R. {Mould}, P.~A. {Scowen}, A.~M.
  {Watson}, J.~A. {Westphal}, {Mapping Jupiter's Latitudinal Bands and Great
  Red Spot Using HST/WFPC2 Far-Ultraviolet Imaging}, Icarus 143 (2000)
  189--204.
\newblock \href {http://dx.doi.org/10.1006/icar.1999.6232}
  {\path{doi:10.1006/icar.1999.6232}}.

\bibitem{Bonfond_10}
B.~{Bonfond}, {Morphology and dynamics of the Io UV footprint}, Ph.D. thesis,
  Universit{\'e} de Li{\`e}ge, Acad{\'e}mie Wallonie-Europe (2010).

\bibitem{Li_IEEE_04}
X.~{Li}, R.~{Ramachandran}, S.~{Movva}, S.~{Graves}, G.~{Germany},
  W.~{Lyatsky}, A.~{Tan}, {Dayglow Removal from FUV Auroral Images},
  Proceedings of the Geoscience and Remote Sensing Symposium 6 (2004)
  3774--3777.
\newblock \href {http://dx.doi.org/10.1109/IGARSS.2004.1369944}
  {\path{doi:10.1109/IGARSS.2004.1369944}}.

\bibitem{Gerard_JGR_04}
J.-C. {G{\'e}rard}, D.~{Grodent}, J.~{Gustin}, A.~{Saglam}, J.~T. {Clarke},
  J.~T. {Trauger}, {Characteristics of Saturn's FUV aurora observed with the
  Space Telescope Imaging Spectrograph}, Journal of Geophysical Research (Space
  Physics) 109 (2004) 9207.
\newblock \href {http://dx.doi.org/10.1029/2004JA010513}
  {\path{doi:10.1029/2004JA010513}}.

\bibitem{Gerard_JGR_06}
J.-C. {G{\'e}rard}, D.~{Grodent}, S.~W.~H. {Cowley}, D.~G. {Mitchell}, W.~S.
  {Kurth}, J.~T. {Clarke}, E.~J. {Bunce}, J.~D. {Nichols}, M.~K. {Dougherty},
  F.~J. {Crary}, A.~J. {Coates}, {Saturn's auroral morphology and activity
  during quiet magnetospheric conditions}, Journal of Geophysical Research
  (Space Physics) 111~(A10) (2006) 12210.
\newblock \href {http://dx.doi.org/10.1029/2006JA011965}
  {\path{doi:10.1029/2006JA011965}}.

\bibitem{Grodent_JGR_05}
D.~{Grodent}, J.-C. {G{\'e}Rard}, S.~W.~H. {Cowley}, E.~J. {Bunce}, J.~T.
  {Clarke}, {Variable morphology of Saturn's southern ultraviolet aurora},
  Journal of Geophysical Research (Space Physics) 110 (2005) 7215.
\newblock \href {http://dx.doi.org/10.1029/2004JA010983}
  {\path{doi:10.1029/2004JA010983}}.

\bibitem{Clarke_JGR_09}
J.~T. {Clarke}, J.~{Nichols}, J.-C. {G{\'e}rard}, D.~{Grodent}, K.~C. {Hansen},
  W.~{Kurth}, G.~R. {Gladstone}, J.~{Duval}, S.~{Wannawichian}, E.~{Bunce},
  S.~W.~H. {Cowley}, F.~{Crary}, M.~{Dougherty}, L.~{Lamy}, D.~{Mitchell},
  W.~{Pryor}, K.~{Retherford}, T.~{Stallard}, B.~{Zieger}, P.~{Zarka},
  B.~{Cecconi}, {Response of Jupiter's and Saturn's auroral activity to the
  solar wind}, Journal of Geophysical Research (Space Physics) 114 (2009) 5210.
\newblock \href {http://dx.doi.org/10.1029/2008JA013694}
  {\path{doi:10.1029/2008JA013694}}.

\bibitem{Lamy_08}
L.~{Lamy}, {Etude des {\'e}missions radio aurorales de Saturne,
  mod{\'e}lisation et aurores UV}, Ph.D. thesis, Universit{\'e} Paris VI
  (2008).

\bibitem{Gustin_JGR_12}
J.~{Gustin}, B.~{Bonfond}, D.~{Grodent}, J.-C. {G{\'e}rard}, {Conversion from
  HST ACS and STIS auroral counts into brightness, precipitated power, and
  radiated power for H$_{2}$ giant planets}, Journal of Geophysical Research
  (Space Physics) 117 (2012) 7316.
\newblock \href {http://dx.doi.org/10.1029/2012JA017607}
  {\path{doi:10.1029/2012JA017607}}.

\bibitem{Lamy_JGR_13}
L.~{Lamy}, R.~{Prang{\'e}}, W.~{Pryor}, J.~{Gustin}, S.~V. {Badman},
  H.~{Melin}, T.~{Stallard}, D.-G. {Mitchell}, P.~C. {Brandt}, {Multispectral
  simultaneous diagnosis of Saturn's aurorae throughout a planetary rotation},
  Journal of Geophysical Research (Space Physics) 118 (2013) 4817--4843.
\newblock \href {http://arxiv.org/abs/1307.4675} {\path{arXiv:1307.4675}},
  \href {http://dx.doi.org/10.1002/jgra.50404} {\path{doi:10.1002/jgra.50404}}.

\bibitem{Pryor_Nature_11}
W.~R. {Pryor}, A.~M. {Rymer}, D.~G. {Mitchell}, T.~W. {Hill}, D.~T. {Young},
  J.~{Saur}, G.~H. {Jones}, S.~{Jacobsen}, S.~W.~H. {Cowley}, B.~H. {Mauk},
  A.~J. {Coates}, J.~{Gustin}, D.~{Grodent}, J.-C. {G{\'e}rard}, L.~{Lamy},
  J.~D. {Nichols}, S.~M. {Krimigis}, L.~W. {Esposito}, M.~K. {Dougherty}, A.~J.
  {Jouchoux}, A.~I.~F. {Stewart}, W.~E. {McClintock}, G.~M. {Holsclaw}, J.~M.
  {Ajello}, J.~E. {Colwell}, A.~R. {Hendrix}, F.~J. {Crary}, J.~T. {Clarke},
  X.~{Zhou}, {The auroral footprint of Enceladus on Saturn}, Nature 472 (2011)
  331--333.
\newblock \href {http://dx.doi.org/10.1038/nature09928}
  {\path{doi:10.1038/nature09928}}.

\bibitem{Wannawichian_JGR_08}
S.~{Wannawichian}, J.~T. {Clarke}, D.~H. {Pontius}, {Interaction evidence
  between Enceladus' atmosphere and Saturn's magnetosphere}, Journal of
  Geophysical Research (Space Physics) 113 (2008) 7217.
\newblock \href {http://dx.doi.org/10.1029/2007JA012899}
  {\path{doi:10.1029/2007JA012899}}.

\bibitem{Erard_AC_14}
S.~{Erard}, P.~{Le Sidaner}, B.~{Cecconi}, J.~{Berthier}, F.~{Henry},
  M.~{Molinaro}, M.~{Giardino}, N.~{Bourrel}, N.~{Andr{\'e}}, M.~{Gangloff},
  C.~{Jacquey}, F.~{Topf}, {The EPN-TAP protocol for the Planetary Science
  Virtual Observatory}, Astronomy and Computing (2014) in press\href
  {http://arxiv.org/abs/1407.5738} {\path{arXiv:1407.5738}}.

\bibitem{Erard_AC_14b}
S.~{Erard}, B.~{Cecconi}, P.~{Le Sidaner}, J.~{Berthier}, F.~{Henry},
  C.~{Chauvin}, N.~{Andr{\'e}}, V.~{G{\'enot}}, C.~{Jacquey}, M.~{Gangloff},
  N.~{Bourrel}, B.~{Schmitt}, M.~T. {Capria}, G.~{Chanteur}, {Planetary Science
  Virtual Observatory architecture}, Astronomy and Computing (2014) in
  press\href {http://arxiv.org/abs/1407.4886} {\path{arXiv:1407.4886}}, \href
  {http://dx.doi.org/10.1016/j.ascom.2014.07.005}
  {\path{doi:10.1016/j.ascom.2014.07.005}}.

\bibitem{Genot_AC_14}
V.~{G{\'enot}}, N.~{Andr{\'e}}, B.~{Cecconi}, M.~{Bouchemit}, E.~{Budnik},
  N.~{Bourrel}, M.~{Gangloff}, N.~{Dufourg}, S.~{Hess}, R.~{Modolo},
  B.~{Renard}, N.~{Lormant}, L.~{Beigbeder}, D.~{Popescu}, J.-P. {Toniutti},
  {Joining the yellow hub: uses of the Simple Application Messaging Protocol in
  Space Physics analysis tools}, Astronomy and Computing (2014) in press\href
  {http://arxiv.org/abs/hal-01055883} {\path{arXiv:hal-01055883}}, \href
  {http://dx.doi.org/10.1016/j.ascom.2014.07.007}
  {\path{doi:10.1016/j.ascom.2014.07.007}}.

\end{thebibliography}

\end{document}